\documentclass[prd,aps,nofootinbib]{revtex4-2}
\usepackage{graphicx}
\usepackage{amsmath,amssymb,bm}
\usepackage{autobreak,mathtools}
\usepackage{comment}
\usepackage[dvipsnames, usenames]{xcolor}
\usepackage{mathrsfs}
\usepackage{hyperref}

\definecolor{p-r}{RGB}{171, 40, 52}
\hypersetup{colorlinks=true, citecolor=p-r, linkcolor=p-r,urlcolor=p-r}
\definecolor{rRGB}{RGB}{169.2, 155.0, 0.5}

\begin{document}

\title{Semiclassical regularity of compact trapped regions:\\ From dynamical horizons to inner extremality}

\author{Ra\'ul Carballo-Rubio}
\affiliation{Instituto de Astrof\'isica de Andaluc\'ia (IAA-CSIC),
Glorieta de la Astronom\'ia, 18008 Granada, Spain}
\author{Francesco Di Filippo}
\affiliation{Institut f\"ur Theoretische Physik, Max-von-Laue-Str. 1, 60438 Frankfurt, Germany}
\author{Stefano Liberati}
\affiliation{SISSA - International School for Advanced Studies, Via Bonomea 265, 34136 Trieste, Italy}
\affiliation{
IFPU - Institute for Fundamental Physics of the Universe, Via Beirut 2, 34014 Trieste, Italy}
\affiliation{INFN Sezione di Trieste, Via Valerio 2, 34127 Trieste, Italy}
\author{Matt Visser}
\affiliation{School of Mathematics and Statistics, Victoria University of Wellington, PO Box 600, Wellington 6140, New Zealand}

\begin{abstract}
In eternal black-hole spacetimes, inner horizons are Cauchy horizons and are generically unstable. For non-extremal inner horizons, this includes both the classical mass-inflation instability and a semiclassical instability associated with divergences in the renormalized stress-energy tensor (RSET). Inner-extremal geometries, for which the inner-horizon surface gravity vanishes, evade classical mass inflation, but in stationary settings still suffer from singular behavior of the RSET. In this work, we show that the dynamical case is qualitatively different. Considering spacetimes describing the formation and evaporation of a compact trapped region in finite time, and working in the $s$-wave Polyakov approximation, we compute the expectation value of the stress-energy tensor in the \emph{in}-vacuum state. 
Given that in this case the inner horizon is \emph{not} a Cauchy horizon, the RSET remains finite everywhere. For generic non-extremal inner horizons, however, the RSET grows exponentially in time at the inner horizon, with a divergence emerging only in the asymptotic limit of an ever-lasting trapped region. For inner-extremal geometries this exponential growth is replaced by a considerably milder power-law growth. Such spacetimes may therefore be considered natural candidates for classically and semiclassically meta-stable black-hole interiors.

\bigskip
\noindent
{\sc Keywords:} regularity, inner horizons, Cauchy horizons, trapping horizons, renormalized stress-energy tensor, Polyakov approximation, semiclassical gravity

\end{abstract}

\maketitle

\clearpage

\tableofcontents

\clearpage
%==========================
\section{Introduction}
%==========================

Black-hole physics has traditionally focused on exterior regions and event horizons~\cite{Carballo-Rubio:2025ghn}. For non-extremal solutions, outer horizons are classically stable under perturbations, while semiclassically they undergo the slow area decrease associated with Hawking evaporation. Although conceptually important, this effect is expected to have limited astrophysical relevance, except perhaps for primordial black holes.

Inner horizons, by contrast, have only recently moved to the forefront. This renewed attention is driven in part by the study of quantum-gravity-inspired regular black holes and their possible observational viability (see e.g.~\cite{Carballo-Rubio:2025fnc} for a review), but the issue is more general --- understanding the stability of black-hole interiors, whether singular or regular, remains a central open problem because inner horizons are not peculiar to regular black holes; they also arise in astrophysically relevant Kerr geometries.

At the classical level, non-extremal inner horizons are well known to suffer from mass inflation~\cite{Simpson:1973ua,Poisson:1989zz,Poisson:1990eh,Bonanno:1994ma,Bonanno:1994qh,Carballo-Rubio:2018pmi,Carballo-Rubio:2021bpr,DiFilippo:2022qkl}, and this instability has recently been shown to also persist for slowly evolving inner horizons~\cite{Carballo-Rubio:2024dca} (see also reference~\cite{Visser:2024zkx}). In physical terms, mass inflation results from the infinite blueshift of outgoing null rays near the inner horizon, which is converted (through interactions with ingoing perturbations/fluxes) into an exponentially growing local energy density. The characteristic timescale is set by $1/|\kappa_-|$, with $\kappa_-$ the inner-horizon surface gravity. 

This immediately suggests a route to evade the instability: geometries with multiply degenerate inner horizons, for which $\kappa_-=0$~\cite{Carballo-Rubio:2022kad} (which can also be defined in the presence of rotation~\cite{Franzin:2022wai}). Such inner-extremal configurations retain an extended trapped region and a non-extremal outer horizon, while being explicitly stable against mass inflation~\cite{Carballo-Rubio:2022kad}.\footnote{Analogous constructions can also be implemented at the outer horizon, leading to geometries with trapped regions and fully extremal horizons; because of their expected stability, these configurations were dubbed ``black-hole graveyards'' in~\cite{DiFilippo:2024spj}. Noticeably, black holes with multi-degenerate horizons were found to be meta-stable (or stable, in the infinite degenerate limit) under the so called Aretakis instability typical of extremal horizons~\cite{Agrawal:2026oka}.}

However, classical stability does not guarantee semiclassical stability. A distinct instability of inner horizons emerges in the renormalized stress-energy tensor (RSET) of quantum fields~\cite{Hollands:2019whz}. In stationary spacetimes, this can be traced to the fact that it is impossible to construct a state regular at both the event and inner horizons whenever their surface gravities differ in magnitude~\cite{Balbinot:2023vcm,McMaken:2023uue,McMaken:2024fvq}. As a result, even black holes that evade (classical) mass inflation through an inner-extremal horizon are not automatically free of semiclassical pathologies~\cite{Barcelo:2020mjw,Barcelo:2022gii,McMaken:2023uue,McMaken:2024fvq}. We emphasize that the notion of ``regularity'' (absence of curvature singularities) is quite distinct from the notion of ``stability'' (the existence of an equilibrium). One of these concepts does not necessarily imply the other.

As explained above, while the extension of mass inflation to slowly evolving geometries is now understood~\cite{Carballo-Rubio:2024dca}, the corresponding question for semiclassical perturbations has not, to our knowledge, been systematically addressed. Even within the framework provided by the Einstein field equations, semiclassical effects might also play a significant role in the structure of the black hole interior and the presence of spacetime singularities~\cite{DiFilippo:2025kzh} (see also~\cite{Gralla:2025gzl}), reinforcing the need for a better understanding of this question in Einstein's theory. In this paper, we study the renormalized expectation value of the stress-energy tensor (RSET) in dynamical spacetimes with an inner horizon, focusing on situations without a Cauchy horizon both within and beyond general relativity.

We shall work in a simplified but physically representative setting: spherically symmetric spacetimes in the $s$-wave Polyakov approximation, which reduces the problem to an effective two-dimensional analysis~\cite{Parentani:1994ij,Cruz:1996ep,Ayal:1997ab,Fabbri:2005zn,Barcelo:2011bb,McMaken:2023uue,Barenboim:2024dko,Barenboim:2025ckx}. Within this framework, we first recover and extend the known stationary results for a set of representative geometries. We then show that, generically, the RSET in the \emph{in}-state --- the state naturally selected by the formation and finite lifetime of a compact trapped region --- remains finite everywhere, but grows in time at the inner horizon. A divergence appears only in the asymptotic limit of an asymptotically stationary trapped region. By contrast, this growth and the associated backreaction are absent for inner-extremal, multi-degenerate horizons, suggesting that such configurations may provide a completely stable (modulo the Hawking evaporation of the outer horizon) endpoint of the evolution.

The paper is organized as follows. In Sec.~\ref{sec:eternal} we review the construction of the relevant quantum states and the computation of the RSET in the Polyakov approximation for eternal spacetimes with Cauchy horizons. The associated regularity conditions are discussed in detail in Sec.~\ref{sec:regularity}. Explicit results for Reissner--Nordstr\"om, Hayward, and inner-extremal black holes are presented in Secs.~\ref{sec:eternalRN}, \ref{sec:eternalH}, and \ref{sec:eternalIE}. Dynamical spacetimes without Cauchy horizons are then analyzed in Sec.~\ref{sec:dynamical}, again for the Reissner--Nordstr\"om, Hayward, and inner-extremal cases, in Secs.~\ref{sec:dynamicalRN}, \ref{sec:dynamicalH}, and \ref{sec:dynamicalIE}. We summarize our conclusions in Sec.~\ref{sec:conclusions}.

%------------------------------------------
\section{Eternal geometries}\label{sec:eternal}
%------------------------------------------

Let us start with a brief recap of known results for eternal geometries. 

%---------------------------
\subsection{Generalities}
%---------------------------

We consider a static spacetime whose metric is given by
\begin{equation}
\label{eq:mettr}
    ds^2=-f(r)\; dt^2\; +\frac{dr^2}{f(r)}+r^2\,d\Omega^2\,.    
\end{equation}
We can introduce the usual tortoise coordinate 
\begin{equation}
    r^\star=\int \frac{dr}{f(r)},
\end{equation}
as well as the Eddington--Finkelstein null coordinates 
\begin{equation}
\label{eq:tort_RN}
    u=t-r^\star,\qquad v=t+r^\star\,,\qquad\implies\qquad r^\star=\frac{v-u}{2},
\end{equation}
in which the metric takes the form
\begin{equation}
\label{eq:metuv}
    ds^2=-f(r(v-u))\; du\; dv+r(v-u)^2\,d\Omega^2\,.
\end{equation}
The first part of the analysis does not depend on the specific choice of the metric function $f(r(v-u))$, 
so we will specify its form later on. 

We have different possible choices for the quantum vacuum state, which correspond  to the different ways we can split the field into positive and negative energy states. This splitting is not unique as it depends on the selection of a specific choice of time function. One possible choice is the Killing time $t=\left(u+v\right)/2$, leading to the splitting
\begin{equation}\label{eq:Boulware}
    \phi=\sum_\omega e^{-i\omega v}a^L_\omega+e^{-i\omega u}a^R_\omega+h.c.\,,
\end{equation}
where the $L (R)$ suffix indicates that the corresponding modes are left(right)-going and $h.c.$ denotes the Hermitian conjugate terms. The Boulware state $\left|B\right>$ is then defined as the vacuum state which is annihilated by the operators $a^L_\omega$ and $a^R_\omega$ in Eq.~\eqref{eq:Boulware}
\begin{equation}
    a^L_\omega\left|B\right>=a^R_\omega\left|B\right>=0\,.
\end{equation}

The expectation value of the SET in the $s$-wave approximation, (Polyakov approximation), calculated from the norm of the conformal Killing vector $\partial_t$, ($g(\partial_t,\partial_t) = -f$), and particularized to the null coordinates used above, reads~\cite{Barcelo:2011bb}:
\begin{equation}\label{eq:B_SET}
    \begin{array}{l}
    \displaystyle\left<B\right|\hat{T}_{uu}\left|B\right>=\left<B\right|\hat{T}_{vv}\left|B\right>= -\frac{1}{192\pi}\left(f'(r)^2-2f(r)f''(r)\right)\,,\\
    \\
   \displaystyle \left<B\right|\hat{T}_{uv}\left|B\right>= \frac{1}{96\pi}f(r)f''(r)\,,
\end{array}
\end{equation}
with all other components zero.

We can now construct different quantum vacuum states by splitting the positive and negative energy modes differently with respect to a different choice of ``preferred'' observers or, from a geometric perspective, with respect to any other conformal Killing vector. Then
\begin{equation}\label{eq:vacuum_generic}
    \phi=\sum_\omega e^{-i\omega \tilde{v}(v)}\tilde a^L_\omega+e^{-i\omega \tilde{u}(u)} \tilde a^R_\omega+h.c.\,,
\end{equation}
where we define the vacuum state $\left|\tilde{0}\right>$ as the state annihilated by the new destruction operators
\begin{equation}
\tilde a^L_\omega\left|\tilde 0\right>=\tilde a^R_\omega\left|\tilde 0\right>=0\,.
\end{equation}
The expectation value of the stress energy tensor in a generic state can be related to the expectation value in the Boulware state by~\cite{Christensen:1977jc,Davies:1977pvx,Fabbri:2005mw}
\begin{equation}\label{eq:SET_generic}
    \begin{array}{l}      
    \displaystyle \left<\tilde 0\right|\hat{T}_{vv}\left|\tilde 0\right>=  \left<B\right|\hat{T}_{vv}\left|B\right>-\frac{1}{24 \pi}\left\{\tilde v(v),v \right\}\,, 
   \\
    \\
   \displaystyle \left<\tilde 0\right|\hat{T}_{uu}\left|\tilde 0\right>=  \left<B\right|\hat{T}_{uu}\left|B\right>-\frac{1}{24 \pi}\left\{\tilde u(u),u \right\}\,, 
   \\ \\
   \displaystyle \left<\tilde 0\right|\hat{T}_{uv}\left|\tilde 0\right>=\left<B\right|\hat{T}_{uv}\left|B\right>\,,
\end{array}
\end{equation}
where we are using the standard definition of the Schwarzian derivative
\begin{equation}\label{eq:def_Schwarzian}
\{y(x),x\}=\frac{y'''(x)}{y'(x)}-\frac{3}{2}\left(\frac{y''(x)}{y'(x)}\right)^2.    
\end{equation}
Throughout the paper we will be using the notation $y'(x)=dy/dx$ for the derivative of a single-variable function with respect to its argument.

To simplify our treatment, we shall assume $f(r)$ to be a rational function of the radial coordinate,
\begin{equation}
f(r)=\frac{P_n(r)}{Q_n(r)},    
\end{equation}
where $P_n(r)$ and $Q_n(r)$ are polynomials of the same degree $n$.  Albeit somewhat restrictive, this assumption can be shown to include all of the most relevant geometries for our investigation.

For concreteness, some specific examples are (see also reference~\cite{Frolov:2016pav}):
\begin{itemize}
    \item \textit{Schwarzschild black hole:} $n=1$, $P_1(r)=r-r_+$, $Q_1(r)=r$.
    \item
    \textit{Reissner--Nordstr\"om black hole:} $n=2$, $P_2(r)=(r-r_+)(r-r_-)$, $Q_2(r)=r^2$, \\
    with $r_\pm=M\pm\sqrt{M^2-Q^2}$.
    \item \textit{Hayward black hole~\cite{Hayward:2005gi}:} $n=3$, $P_3(r)=(r-r_-)(r-r_+)\left[(r_++r_-)r+r_+r_-\right]$, \\
    $Q_3(r)=(r_++r_-)r^3+r_+^2r_-^2$, with $r_\pm$ implicitly determined by the equations \\
    $\ell=r^+r^-/\sqrt{r_+^2+r_+r_-+r_-^2}$ and $M=\left(r_+^2+r_+r_-+r_-^2\right)/\left[2(r_++r_-)\right]$. See~\cite{Frolov:2016pav}.
    \item \textit{Inner-extremal black hole:} $n=4$, $P_4(r)=(r-r_+)(r-r_-)^3$, whereas \\
    $Q_4(r)=\left(r - r_-\right)^3 \left(r - r_+\right) + 2 M r^3 + b r^2$, with $b$ large enough to avoid zeros in the denominator. See~\cite{Carballo-Rubio:2022kad}.
\end{itemize}
We will follow the convention that $r_+$ indicates the position of the outer horizon and $r_-$ the position of the inner horizon (in the sense of Hayward~\cite{Hayward:1993wb}). In stationary situations, these also correspond to event and Cauchy horizons, respectively.

%--------------------------------------------
\subsection{Regularity of a quantum state}\label{sec:regularity}
%--------------------------------------------

We recall that a quantum state is regular if the expectation value of the stress energy tensor is regular in a well-behaved set of coordinates. Thus, the expressions in Eq.~\eqref{eq:SET_generic} are insufficient to study the regularity of the expectation values of the stress energy tensor whenever $f(r)$ has roots, as the null coordinates $(u,v)$ become ill-defined there due to the determinant of the metric defined in Eq.~\eqref{eq:metuv} vanishing. However, considering situations in which $f(r)$ has roots is necessary in order to describe different black hole spacetimes, and, so we need to introduce Kruskal coordinates, regular at both the event and Cauchy horizons. 

The precise way to implement Kruskal coordinates depends on the spacetime under consideration, and in particular let us recall that they require the identification of the surface gravities associated to the horizons. In what follows, we use the standard time-dependent definition of the surface gravity encapsulating the peeling of null rays around outer and inner horizons~\cite{Cropp:2013zxi}, i.e.
\begin{equation}\label{eq:surface_grav}
    \kappa_\pm=\left.\frac{1}{2}\frac{df}{d r}\right|_{r=r_\pm}\,.
\end{equation}
The actual expression for the Kruskal coordinates depends on whether the surface gravity vanishes or not. In the following we consider first the case in which neither surface gravity vanishes,  and subsequently the case in which $\kappa_-=0$ while $\kappa_+\neq 0$. Other possible combinations can be studied in the same way, but are not the primary focus of this investigation.

%--------------------------------------------
\subsection{Reissner--Nordstr\"om spacetime}\label{sec:eternalRN}
%--------------------------------------------
The first case we consider is the non-extremal Reissner--Nordstr\"om geometry, whose semiclassical properties were also studied in \cite{Balbinot:2023vcm}. While this is a special case, the discussion remains qualitatively the same for any other geometry with non-extremal horizons, including the Hayward or Bardeen regular black holes, with expressions differing just in a few numerical factors. We will show this explicitly for the Hayward black hole in the section below.

\begin{figure}[!h]
    \centering
    \includegraphics[width=0.4\linewidth]{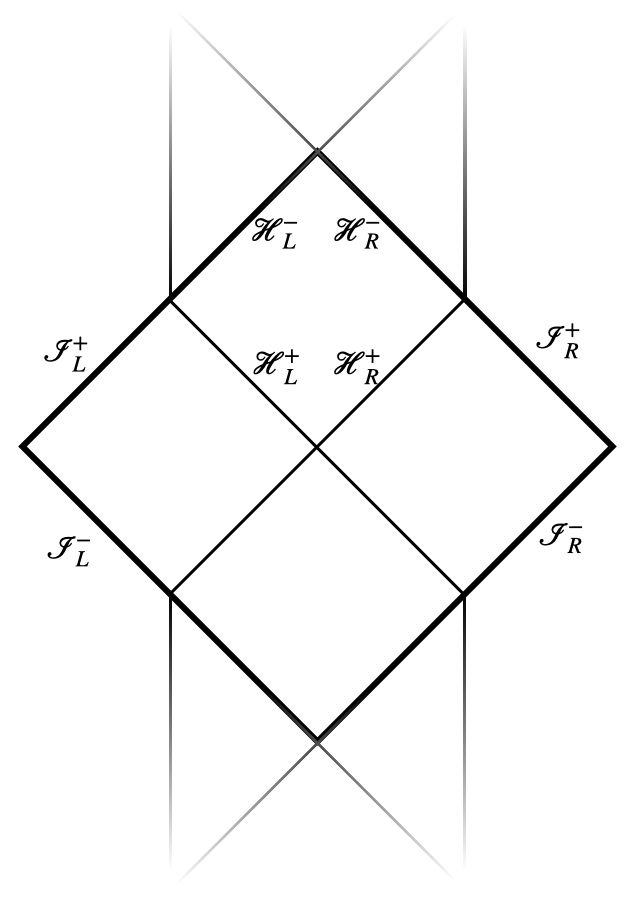}
    \caption{(Partial) conformal diagram of an eternal black hole with two horizons, including the domains of outer communication and down to the (future) inner horizon. For the purposes of the current discussion we will mainly focus attention on the diamond shaped region between the future outer and future inner horizons.}
    \label{fig:eternal_sptm}
\end{figure}

For the reader's convenience, we show in Fig.~\ref{fig:eternal_sptm} the causal structure of the relevant sectors of the spacetime.

\subsubsection{Boulware state}

The expectation value of the stress energy tensor in the Boulware state \eqref{eq:Boulware} can be easily obtained from Eq.~\eqref{eq:B_SET} and reads
\begin{align}\label{eq:B_SET_RN}
\displaystyle\left<B\right|\hat{T}_{uu}\left|B\right>&=\frac{-4 r^3 (r_-+r_+)+3 r^2 \left(r_-^2+6 r_-
   r_++r_+^2\right)-12 r r_- r_+
   (r_-+r_+)+8 r_-^2 r_+^2}{48\pi r^3}\,,\\
\left<B\right|\hat{T}_{vv}\left|B\right>&=\left<B\right|\hat{T}_{uu}\left|B\right>, \\
   \displaystyle \left<B\right|\hat{T}_{uv}\left|B\right>&= \frac{1}{48\pi r^6}(r-r_-) (r-r_+) \left[3 r_- r_+-r
   (r_-+r_+)\right]\,.
\end{align}
While these expectation values are finite everywhere, we still cannot conclude that the stress energy tensor is well defined everywhere as the coordinate system used does not cover the horizons. In particular, we have\footnote{The signs of the limits are fixed by the behaviour of the tortoise coordinate, $r^\star=(v-u)/2$. Near the outer horizon this can be expanded as (see also the discussion below)
\[
r^\star\sim {1\over 2\kappa_+}\log|r-r_+|\to -\infty,
\qquad \kappa_+>0,
\]
whereas near the inner horizon,
\[
r^\star\sim {1\over 2\kappa_-}\log|r-r_-|\to +\infty,
\qquad \kappa_-<0.
\] So, in the trapped region, where $r^\star$ is the timelike coordinate, it runs from $-\infty$ at the outer horizon, up to $+\infty$ at the inner horizon.}
\begin{equation}\label{eq_EF_horizon}
\begin{array}{c} 
    u\rightarrow +\infty\qquad \text{at} \qquad \mathscr{H}^+_R\,;\qquad\qquad \qquad 
    u\rightarrow -\infty\qquad \text{at} \qquad \mathscr{H}^-_L\,; \\
    v\rightarrow -\infty\qquad \text{at} \qquad \mathscr{H}^+_L\,;\qquad \qquad\qquad 
    v\rightarrow +\infty\qquad \text{at} \qquad \mathscr{H}^-_R\,.
\end{array}
\end{equation}

We can then introduce a Kruskal coordinate $U_-$ well defined at the left sector of the inner horizon, see Fig.~\ref{fig:eternal_sptm}, $\mathscr{H}^-_L$, in terms of which the metric reads
\begin{equation}
        ds^2=-\left[f\left(r(U_-,v)\right)\;\frac{du}{dU_-}\right]\; dU_- \; dv+r^2(u,v)d\Omega^2\,,
\end{equation}
Regularity of this metric at the inner horizon requires that the quantity
\begin{equation}
    f(r(U_-,v))\; \frac{du}{dU_-}
\end{equation}
is both finite and non-vanishing there in spite of the fact that such location is a root for $f(r)$. 

We hence need to determine how rapidly the function $f(r(u,v))$ vanishes in the vicinity of the inner horizon. To this end, we note that the tortoise coordinate is given by
\begin{equation}\label{eq:tortoiseRN}
        r^\star=\int \frac{dr}{f(r)}
        =r+\frac{1}{2\kappa_-}\log\left|\kappa_-(r-r_-)\right|+\frac{1}{2\kappa_+}\log\left|\kappa_+(r-r_+)\right|\,,
\end{equation}
where we have used Eq.~\eqref{eq:surface_grav} to write
\begin{equation}
\kappa_+=\frac{r_+-r_-}{2r_+^2}\,,\qquad \kappa_-=\frac{r_--r_+}{2r_-^2}\,.
\label{eq:RNkappas}
\end{equation}

Near the inner horizon, $r=r_-$, the only singular contribution in Eq.~\eqref{eq:tortoiseRN} is the logarithmic term involving $(r-r_-)$, while all the remaining terms stay finite. Therefore,
\begin{equation}
r^\star = C_- + \frac{1}{2\kappa_-}\log\left|\kappa_-(r-r_-)\right| + \mathcal O(r-r_-)\,,
\end{equation}
where $C_-$ is a finite constant. It follows that
\begin{equation}
\log\left|\kappa_-(r-r_-)\right| = 2\kappa_-\bigl(r^\star-C_-\bigr)+\mathcal O(r-r_-)\,,
\end{equation}
and hence
\begin{equation}\label{eq:u_r}
r-r_- \propto e^{2\kappa_- r^\star}=e^{\kappa_-(v-u)}\,.
\end{equation}
In particular, approaching $\mathscr{H}^-_L$ at fixed $v$ and $u\to -\infty$, one has
\begin{equation}
r-r_- \propto e^{-\kappa_- u}\,.
\end{equation}

Since near $r=r_-$ one also has
\begin{equation}
f(r)=f'(r_-)(r-r_-)+\mathcal O\!\left((r-r_-)^2\right)=2\kappa_-(r-r_-)+\mathcal O\!\left((r-r_-)^2\right)\,,
\end{equation}
we conclude that, in the vicinity of $\mathscr{H}^-_L$,
\begin{equation}
f(r(u,v))\propto e^{-\kappa_- u}\,.
\end{equation}
Therefore, regularity of the metric requires
\begin{equation}\label{eq:U_-innerextr_RN}
\frac{du}{dU_-}\propto e^{\kappa_- u}
\qquad\iff\qquad
U_-\propto e^{-\kappa_- u}+ \mathrm{const.}
\end{equation}

In a similar way, we can also introduce a coordinate $V_-$ which is well defined at the right sector of the inner horizon,~$\mathscr{H}^-_R$, and the usual Kruskal coordinates $U_+$ and $V_+$, respectively regular at the right and left outer horizons.
Suitably fixing the free proportionality constants, we obtain
\begin{equation}\label{eq:Kruskal_-}
    U_-= -\frac{1}{\kappa_-}e^{-\kappa_-u}=\frac{1}{|\kappa_-|}e^{|\kappa_-|u}\,,\qquad V_-= \frac{1}{\kappa_-}e^{\kappa_-v}=-\frac{1}{|\kappa_-|}e^{-|\kappa_-|v}\,,
\end{equation}
and
\begin{equation}\label{eq:Kruskal_+}
    U_+= -\frac{1}{\kappa_+}e^{-\kappa_+u}\,,\qquad V_+= \frac{1}{\kappa_+}e^{\kappa_+v}\,.
\end{equation}
Please note that, given our definitions~\eqref{eq:RNkappas}, $\kappa_-<0$ and $\kappa_+>0$ in the above expressions. Also note that the above definitions refer to the Kruskal coordinates respectively regular on the \emph{exterior} of the four horizons of our interest. Extension of these coordinates into the \emph{interiors} requires one to apply an overall minus sign, which however will be irrelevant in the RSET expectation values as only squared quantities will appear there. The same applies for all the other Kruskal coordinates we shall define in the discussion below.

It is straightforward to check that the metric is explicitly regular at the outer horizons when expressed in terms of $(U_+,V_+)$ coordinates, and at the inner horizons when expressed in terms of $(U_-,V_-)$ coordinates. 
Furthermore, from the definitions of the Kruskal coordinates and the behavior of the null coordinates in Eq.~\eqref{eq_EF_horizon}, we infer that the corresponding Kruskal coordinate vanishes on each horizon:
\begin{equation}\label{eq_Kruskal_horizon}
\begin{array}{c} 
    U_+\sim (r-r_+)=0\qquad \text{at} \qquad \mathscr{H}^+_R\,;\qquad\qquad  
    U_-\sim (r-r_-)=0\qquad \text{at} \qquad \mathscr{H}^-_L\,; \\
    V_+\sim (r-r_+)=0\qquad \text{at} \qquad \mathscr{H}^+_L\,;\qquad \qquad 
     V_-\sim (r-r_-)=0\qquad \text{at} \qquad \mathscr{H}^-_R\,.
\end{array}
\end{equation}

The expectation values of Eq.~\eqref{eq:B_SET_RN} expressed in these Kruskal coordinates read
\begin{subequations}\label{eq:B_state_Kruskal}
\begin{align}
     \left<B\right|\hat{T}_{U_\pm U_\pm}\left|B\right>&=\left(\frac{d u}{d U_\pm}\right)^2 \left<B\right|\hat{T}_{uu}\left|B\right>=\kappa_\pm^{-2}U_\pm^{-2}\left<B\right|\hat{T}_{uu}\left|B\right>\,,
     \\
    \left<B\right|\hat{T}_{V_\pm V_\pm}\left|B\right>&=\left(\frac{d v}{d V_\pm}\right)^2 \left<B\right|\hat{T}_{vv}\left|B\right>=\kappa_\pm^{-2}V_\pm^{-2}\left<B\right|\hat{T}_{vv}\left|B\right>\,,
    \\
    \left<B\right|\hat{T}_{U_\pm V_\pm}\left|B\right>&=\frac{d u}{d U_\pm}\frac{d v}{d V_\pm} \left<B\right|\hat{T}_{uv}\left|B\right>=\kappa_\pm^{-2}U_\pm^{-1} V_\pm^{-1}\left<B\right|\hat{T}_{uv}\left|B\right>\,.
     \end{align}
\end{subequations}
As at the horizons we have either $U_\pm=0$ or $V_\pm=0$, Eq.~\eqref{eq:B_state_Kruskal} shows that some components of the stress energy tensor in the Boulware state diverge at either horizon\footnote{Note that the $ \left<B\right|\hat{T}_{U_\pm V_\pm}\left|B\right>$ component is always finite as $T_{uv}$ vanishes at the horizons.}. The behavior of the renormalized stress-energy tensor in the Boulware state is summarized in Table~\ref{tab:rsetcompsB}.

\subsubsection{Unruh state (outer)}

We can define vacuum states that are regular at some of the horizons by splitting the positive and negative energy modes in a way different from Eq.~\eqref{eq:Boulware}. For instance, we can define the Unruh state $\left|U
_+\right>$ regular on the outer horizon $\mathscr{H}^+_R$ by splitting the energy modes as 
\begin{equation}\label{eq:Unruh_+}
    \phi=\sum_\omega e^{-i\omega v}a^L_\omega+e^{-i\omega U_+}a^R_\omega+h.c.\,,
\end{equation}
which correspond to the transformation in Eq.~\eqref{eq:SET_generic} in the case in which $\tilde v=v$ and $\tilde u=U_+$.

Using Eq.~\eqref{eq:SET_generic} we can compute the expectation values in the state $\left|U_+\right>$ 
\begin{equation}
    \left<U_+\right|\hat{T}_{U_+ U_+}\left|U_+\right> = \kappa_+^{-2}U^{-2}_+
\left(\left<B\right|\hat{T}_{uu}\left|B\right>-\frac{1}{24\pi}\left\{U_+,u\right\}\right)\,.
\end{equation}
It is now straightforward to check that 
\begin{equation}
\{U_\pm,u\}= -\frac12 \kappa_\pm^2, \qquad\text{and}\qquad \{V_\pm,v\}= -\frac12 \kappa_\pm^2\,,   
\end{equation}
and thus
\begin{equation}
    \left<U_+\right|\hat{T}_{U_+ U_+}\left|U_+\right>=\kappa_+^{-2}U_+^{-2}\left[\left(\frac{3 r_-^2-4 r_- r_++r_+^2}{32 \pi 
   r_+^6}\right)(r-r_+)^2+\mathcal{O}(r-r_+)^3\right]\,.
\end{equation}
We can thus confirm that this component is regular at $\mathscr{H}^+_R$. On the other hand, the component
\begin{equation}
     \left<U_+\right|\hat{T}_{V_+ V_+}\left|U_+\right>=\kappa_+^{-2}V^{-2}_+
\left<B\right|\hat{T}_{vv}\left|B\right>=-\kappa_+^{-2}V^{-2}_+\left[\frac{(r_--r_+)^2}{192 \pi  r_+^4}+\mathcal{O}(r-r_+)^2\right]\,,
\end{equation}
is divergent at $\mathscr{H}^+_L$. Taking into account the relations in Eq.~\eqref{eq_Kruskal_horizon} the degree of divergence is polynomial as a function of the radial distance from the horizon, but exponential in retarded time
\begin{equation}
    \left<U_+\right|\hat{T}_{V_+ V_+}\left|U_+\right>\propto (r-r_+)^{-2}\propto e^{2\kappa_+ v}\,.
\end{equation}
It is also interesting to compute the expectation values in the state $\left|U_+\right>$ at the inner horizon
\begin{equation}
    \left<U_+\right|\hat{T}_{U_- U_-}\left|U_+\right>=\kappa_-^{-2}U_-^{-2}\left(\left<B\right|\hat{T}_{uu}\left|B\right>-\frac{1}{24\pi}\left\{U_+,u\right\}\right)=\kappa_-^{-2}U_-^{-2}\left[\frac{\kappa_+^2-\kappa_-^2}{48\pi}+\mathcal{O}(r-r_-)^2\right]\,,
\end{equation}
which is clearly divergent on the left branch of the inner horizon, $\mathscr{H}^-_L$ where $U_-=0$.
The other diagonal component reads
\begin{equation}
   \left<U_+\right|\hat{T}_{V_- V_-}\left|U_+\right> = \kappa_-^{-2}V_-^{-2}\left<B\right|\hat{T}_{vv}\left|B\right>=-\kappa_-^{-2}
   V_-^{-2}\left[\frac{\kappa_-^2}{48\pi}+\mathcal{O}(r-r_-)^2\right]\,,
\end{equation}
which implies a divergent behaviour on the right branch of the inner horizon, $\mathscr{H}^-_R$, where $V_-=0$.
Finally, the non-diagonal component,
\begin{equation}
    \left<U_+\right|\hat{T}_{V_- U_-}\left|U_+\right> = \kappa_-^{-2}V_-U_-\left<B\right|\hat{T}_{vu}\left|B\right>=\frac{\kappa_-^{-2}V_-U_-}{48\pi r^6}(r-r_-) (r-r_+) \left[3 r_- r_+-r
   (r_-+r_+)\right]\,,
\end{equation}
is everywhere finite.

The overall behaviour of the stress energy tensor at the outer horizon in the (outer) Unruh state is summarized in Table~\ref{tab:rsetcompsU+}.

\subsubsection{Unruh state (inner)}

Similarly,  we can define an ``inner horizon" Unruh state $\left|U
_-\right>$, which we expect to be regular at $\mathscr{H}^-_L$, by splitting the energy modes as 
\begin{equation}\label{eq:Unruh_-}
    \phi=\sum_\omega e^{-i\omega v}a^L_\omega+e^{-i\omega U_-}a^R_\omega+h.c.\,.
\end{equation}
As before, we can use Eq.~\eqref{eq:SET_generic} to compute the expectation values in the state $\left|U_-\right>$ 
\begin{equation}
    \left<U_-\right|\hat{T}_{U_- U_-}\left|U_-\right> = \kappa_-^{-2}U^{-2}_-
\left(\left<B\right|\hat{T}_{uu}\left|B\right>-\frac{1}{24\pi}\left\{U_-,u\right\}\right)\,.
\end{equation}
We can see that this component
\begin{equation}
    \left<U_-\right|\hat{T}_{U_- U_-}\left|U_-\right>=\kappa_-^{-2}U_-^{-2}\left[\left(\frac{3 r_+^2-4 r_- r_++r_-^2}{32 \pi 
   r_-^6}\right)(r-r_-)^2+\mathcal{O}(r-r_-)^3\right]\,.
\end{equation}
is regular, as expected, at the left branch of the inner horizon (given that $U\sim (r-r_-)$ there), while
\begin{equation}
    \left<U_-\right|\hat{T}_{V_- V_-}\left|U_-\right>=\kappa_-^{-2}V^{-2}_-
\left<B\right|\hat{T}_{vv}\left|B\right>=-\kappa_-^{-2}V^{-2}_-\left[\frac{(r_--r_+)^2}{192 \pi  r_-^4}+\mathcal{O}(r-r_-)^2\right]\,,
\end{equation}
is divergent at the right branch $\mathscr{H}^-_R$. 

Finally, we can compute the expectation value in the state $\left|U_-\right>$  at the outer horizon right branch
\begin{equation}
    \left<U_-\right|\hat{T}_{U_+ U_+}\left|U_-\right>=\kappa_+^{-2}U_+^{-2}\left[\frac{\kappa_-^2-\kappa_+^2}{48\pi}+\mathcal{O}(r-r_+)^2\right]\,,
\end{equation}
which clearly shows a divergence due to the fact that $U_+\to 0$ at $\mathscr{H}^+_R$.
Similar results hold for the left branch of the outer horizon. The expectation value
\begin{equation}
       \left<U_-\right|\hat{T}_{V_+ V_+}\left|U_-\right> = \kappa_+^{-2}V_+^{-2}\left<B\right|\hat{T}_{vv}\left|B\right>=-\kappa_+^{-2}
   V_+^{-2}\left[\frac{\kappa_+^2}{48\pi}+\mathcal{O}(r-r_+)^2\right]\,,
\end{equation}
 $\left<U_-\right|\hat{T}_{V_+ V_+}\left|U_-\right>$ is divergent on $\mathscr{H}^+_L$, in the limit $V_+\to 0$. 
 The overall behaviour of the stress energy tensor in the Unruh state is summarized in Table~\ref{tab:rsetcompsU-}.

%------------------------------------------
\subsection{Hayward spacetime}\label{sec:eternalH}
%------------------------------------------

For the Hayward spacetime, the expectation value of the stress energy tensor can be calculated in close parallel to the calculation for the Reissner--Nordstr\"om spacetime.

\subsubsection{Boulware state}

The RSET in the Boulware state, as calculated using Eq.~\eqref{eq:B_SET}, takes the form
\begin{align}\label{eq:B_SET_RN_H}
\displaystyle\left<B\right|\hat{T}_{uu}\left|B\right>=&-\frac{r_-^2+r_-
   r_++r_+^2}{192 \pi  \left[r^3
   (r_-+r_+)+r_-^2
   r_+^2\right]^4} \left\{4 r^9
   (r_-+r_+)^3-24 r^6 r_-^2
   r_+^2 (r_-+r_+)^2\right.\nonumber\\&
   \left.-3 r^8
   (r_-+r_+)^2
   \left(r_-^2+r_-
   r_++r_+^2\right)-24 r^3 r_-^4
   r_+^4 (r_-+r_+)+4 r_-^6
   r_+^6\right.\nonumber\\
   &\left.+24 r^5
   r_-^2 r_+^2 (r_-+r_+)
   \left(r_-^2+r_-
   r_++r_+^2\right)\right\}\nonumber\,,\\
\left<B\right|\hat{T}_{vv}\left|B\right>=&\left<B\right|\hat{T}_{uu}\left|B\right>, \\
\displaystyle \left<B\right|\hat{T}_{uv}\left|B\right>=&-\frac{(r-r_-) (r-r_+)
   \left(r_-^2+r_-
   r_++r_+^2\right)\left[r
   (r_-+r_+)+r_- r_+\right]}{48 \pi  \left[r^3
   (r_-+r_+)+r_-^2
   r_+^2\right]^4}\left\{r^6 (r_-+r_+)^2\right.\,\nonumber\\
   &\left.-7 r^3
   r_-^2 r_+^2
   (r_-+r_+)+r_-^4
   r_+^4\right\}\,.
\end{align}
The important aspects of these expectations values are that the $(u,u)$ and $(v,v)$ components are non-vanishing at the horizons while the $(u,v)$ component vanishes linearly as $r-r_-$ at the inner horizon, and linearly as $r-r_+$ at the outer horizon.

The definition of regular states at the different branches of the outer and inner horizons follows the same steps as in the previous section. The only difference is that the tortoise coordinate is now
\begin{equation}\label{eq:tortoiseHay}
        r^\star=
        r+\frac{1}{2\kappa_-}\log\left|\kappa_-(r-r_-)\right|
        +\frac{1}{2\kappa_+}\log\left|\kappa_+(r-r_+)\right|
        +\gamma \log \left[\frac{r(r_-+r_+)+r_-r_+}{(r_-+r_+)^2}\right]\,,
\end{equation}
where the last term is regular at both horizons, while the surface gravities are given by 
\begin{equation}
\kappa_+=\frac{(r_+-r_-)(2r_+ + r_-)}
{2\,r_+\,(r_+^2+r_+r_-+r_-^2)}\,,
\qquad
\kappa_-=\frac{(r_- - r_+)(r_+ + 2r_-)}
{2\,r_-\,(r_+^2+r_+r_-+r_-^2)}\,.
\end{equation}
Note that $\kappa_+>0$ and $\kappa_-<0$, as expected.

Therefore the Kruskal coordinates are defined as
\begin{equation}\label{eq:Kruskal_-_H}
    U_-= -\frac{1}{\kappa_-}e^{-\kappa_-u}=\frac{1}{|\kappa_-|}e^{|\kappa_-|u}\,,\qquad
    V_-= \frac{1}{\kappa_-}e^{\kappa_-v}=-\frac{1}{|\kappa_-|}e^{-|\kappa_-|v}\,,
\end{equation}
and
\begin{equation}\label{eq:Kruskal_+_H}
    U_+= -\frac{1}{\kappa_+}e^{-\kappa_+u}\,,\qquad
    V_+= \frac{1}{\kappa_+}e^{\kappa_+v}\,.
\end{equation}

The qualitative behavior of the different quantum states follows the same structure as in Table \ref{tab:rsetcompsB}.  In particular, for the Boulware state, some components of the stress energy tensor always diverge at least at one of the four horizons.

\subsubsection{Unruh states (inner and outer)}

Also in the case of the Unruh states the qualitative behaviour is the same as that shown in Tables \ref{tab:rsetcompsU+} and \ref{tab:rsetcompsU-}. However, the specific functional expressions of the different components of the expectation values of the stress-energy tensor is generally different, although the calculation follows the same steps as before. For completeness, we show some of the components explicitly:
\begin{equation}
    \left<U_-\right|\hat{T}_{U_- U_-}\left|U_-\right> = \kappa_-^{-2}U^{-2}_-
\left(\left<B\right|\hat{T}_{uu}\left|B\right>-\frac{1}{24\pi}\left\{U_-,u\right\}\right)\,.
\end{equation}
Near the inner horizon this reads
\begin{align}\label{eq:Hayward_Uminus}
    \left<U_-\right|\hat{T}_{U_- U_-}\left|U_-\right> &=  \kappa_-^{-2}U^{-2}_-K_-(r_+,r_-)(r-r_-)^2+\mathcal{O}(r-r_-)^3\,,
    \end{align}
with
\begin{equation}
    K_-(r_+,r_-)=\frac{r_-^7+4
  r_-^6 r_+-12 r_-^5
   r_+^2-50 r_-^4 r_+^3-11
   r_-^3 r_+^4+66 r_-^2
   r_+^5+22 r_- r_+^6-20
   r_+^7}{32 \pi  r_-^3
   \left(r_-^2+r_-
   r_++r_+^2\right)^4}\,.
\end{equation}
which is regular at both the left and right branches of the inner horizon. In the former case indeed $U_-\to 0$ as $r-r_-$, while in the latter it stays finite as one reaches $\mathscr{H}^{-}_R$. 

Different is the case of $\left<U_-\right|\hat{T}_{V_- V_-}\left|U_-\right>$ 
\begin{equation}
    \left<U_-\right|\hat{T}_{V_- V_-}\left|U_-\right> =\kappa_-^{-2}V^{-2}_-
\left<B\right|\hat{T}_{vv}\left|B\right>
\end{equation}
which near the inner horizon reads
\begin{equation}
  \left<U_-\right|\hat{T}_{V_- V_-}\left|U_-\right> =  \kappa_-^{-2}V^{-2}_-\left\{-\frac{\left(r_-^2+r_- r_+-2
   r_+^2\right)^2}{192 \left[\pi  r_-^2
   \left(r_-^2+r_-
   r_++r_+^2\right)^2\right]}+\mathcal{O}\left[(r-r_-)^2\right]\right\}\,,
\end{equation}
showing a divergence at $\mathscr{H}^{-}_R$ as there $V_-\to 0$ without any compensation from the numerator.

Finally, the behaviour at the outer horizons is the same as in the RN case, so the RSET is not regular there.
\begin{equation}
\left<U_+\right|\hat{T}_{U_+ U_+}\left|U_+\right> =\kappa_+^{-2}U^{-2}_+K_+(r_+,r_-)(r-r_+)^2+\mathcal{O}(r-r_+)^3\,.
\end{equation}
with
\begin{equation}
    K_+(r_+,r_-)=\frac{ -20 r_-^7+22 r_-^6 r_++66
   r_-^5 r_+^2-11 r_-^4 r_+^3-50 r_-^3
   r_+^4-12 r_-^2 r_+^5+4 r_-
   r_+^6+r_+^7}{32 \pi  r_+^3
   \left(r_-^2+r_-
   r_++r_+^2\right)^4}\,.
\end{equation}
This quantity is regular at $\mathscr{H}^+_R$, while
\begin{equation}
     \left<U_+\right|\hat{T}_{V_+ V_+}\left|U_+\right>=\kappa_+^{-2}V^{-2}_+
\left<B\right|\hat{T}_{vv}\left|B\right>=-\kappa_+^{-2}V^{-2}_+\left[\frac{\kappa_+^2}{48 \pi  
%r_+^4
}+\mathcal{O}(r-r_+)^2\right]\,,
\end{equation}
is divergent at $\mathscr{H}^+_L$. 

%------------------------------------------
\subsection{Inner-extremal spacetime}\label{sec:eternalIE}
%------------------------------------------

Let us now consider the inner-extremal example introduced above, a spacetime with vanishing surface gravity at the inner horizon, while the surface gravity at the outer horizon is non-zero. 

\subsubsection{Boulware state}

The components of the stress-energy in the Boulware state read:
\begin{align}\label{eq:B_SET_RN_H2}
\displaystyle\left<B\right|\hat{T}_{uu}\left|B\right>=&-\frac{(r-r_-)^4}{192 \pi 
   Q_4(r)^4}\left\{\left[Q_4(r) (4
   r-r_--3 r_+)-(r-r_-)
   (r-r_+) Q_4'(r)\right]^2\right.\nonumber\\
   &-2
   (r-r_+)\left[-Q_4(r) (r-r_-)^2
   (r-r_+) Q_4''(r)\right.\nonumber\\
   &+2 (r-r_-)^2
   (r-r_+) Q_4'(r)^2-6 Q_4(r)
   (r-r_-) (r-r_+) Q_4'(r)\nonumber\\
   &-2
   Q_4(r) (r-r_-)^2 Q_4'(r)+6
   Q_4(r)^2 (r-r_-)\left.\left.+6 Q_4(r)^2
   (r-r_+)\right]\right\}\,,\\
\left<B\right|\hat{T}_{vv}\left|B\right>=&\left<B\right|\hat{T}_{uu}\left|B\right>, \\
\displaystyle \left<B\right|\hat{T}_{uv}\left|B\right>=&-\frac{(r-r_-)^4 (r-r_+)}{96 \pi 
   Q_4(r)^4} \left\{-2
   (r-r_-)^2 (r-r_+)
   Q_4'(r)^2\right. \\
   
   &\left.+Q_4(r) (r-r_-)
   \left[(r-r_-) (r-r_+)
   Q_4''(r)+Q_4'(r) (8 r-2r_--6
   r_+)\right]\right.\nonumber \\
   
   &\left.6 Q_4(r)^2 (-2
   r+r_-+r_+)\right\}\nonumber\,.
\end{align}

To define the coordinates regular at the horizons we need the expression of the tortoise coordinate, given by
\begin{equation}\label{eq:tort_Inner_extr}
    r^\star=\int \frac{dr}{f(r)}=r+\frac{A}{(r-r_-)^2}+\frac{B}{(r-r_-)}+C\log \left|\kappa_+(r-r_-)\right|+\frac{1}{2\kappa_+}\log \left|\kappa_+(r-r_+)\right|\,,
\end{equation}
where $A$, $B$, $C$ are suitable constants. In the following, we will only need the two leading terms, controlled by the values of $A$ and $B$, which are equal to
\begin{equation}\label{eq:A-B}
    A=\frac{D(r_-)}{2(r_+-r_-)}\,,\qquad \text{and} \qquad B=\frac{br_- \left(r_--2 r_+\right)+2 M r_-^2 \left(2 r_--3
   r_+\right)}{(r_--r_+)^2}\,.
\end{equation}
For completeness, we also provide the value of $C$ 
\begin{equation}\label{eq:C}
C=\frac{b r_+^2+2 M r_-
   \left(r_-^2-3 r_- r_++3
   r_+^2\right)}{(r_--r_+)^3}.
\end{equation}
Putting together Eq.~\eqref{eq:tort_RN} and Eq.~\eqref{eq:tort_Inner_extr} we obtain
\begin{equation}\label{eq:EF_Inner_extr}
    \frac{v-u}{2}=r+\frac{A}{(r-r_-)^2}+\frac{B}{(r-r_-)}+C\log \left|\kappa_+(r-r_-)\right|+\frac{1}{2\kappa_+}\log \left|\kappa_+(r-r_+)\right|,
\end{equation}
Near the left sector of the inner horizon
\begin{equation}
    u=-\frac{2A}{(r-r_-)^2}+\mathcal{O}(r-r_-)^{-1}\, ,
\end{equation}
while near the right sector of the inner horizon
\begin{equation}
    v=\frac{2A}{(r-r_-)^2}+\mathcal{O}(r-r_-)^{-1}\,.
\end{equation}

We can introduce a coordinate well defined at the sector $\mathscr{H}^-_L$  of the inner horizon, $U_-$. As in the case with non-vanishing surface gravity, we need to impose the condition that the quantity
\begin{equation}
    f(r(U_-,v))\; \frac{du}{dU_-}
\end{equation}
is both finite and non-vanishing. The function $f(r(v-u))$ vanishes cubically as $r$ approaches the inner horizon, while in turn the $u$ coordinate diverges quadratically. Overall we have
\begin{equation}
    f\propto(r-r_-)^3\propto u^{-3/2}\,.
\end{equation}
Therefore, we need to impose 

\begin{equation}\label{eq:U_-innerextr}
    \frac{du}{dU_-}\propto |u|^{3/2}
    \qquad\Longleftrightarrow\qquad
    U_-\propto |u|^{-1/2}+\text{const.}
\end{equation}

In a similar way, we can introduce a coordinate $V_-$ which is well defined at the right sector of the inner horizon
\begin{equation}
    V_-\propto v^{-1/2}+const\,.
\end{equation}
The coordinates regular at the outer horizon have the same expression obtained before for the non-extremal case. Therefore, we obtain
\begin{equation}\label{eq:Kruskal_-_IE}
    U_-=c_1{|u|}^{-1/2}+c_2 \,,\qquad\qquad V_-= \tilde{c}_1{|v|}^{-1/2}+\tilde c_2 ,
\end{equation}
where $c_\textsc{{i}}$ and $\tilde c_\textsc{{i}}$ are free constants that we choose in the following as $c_1=\tilde{c}_1=1$ and $c_2=\tilde{c}_2=0$, while
\begin{equation}\label{eq:Kruskal_+_IE}
    U_+= \mp \frac{1}{\kappa_+}e^{-\kappa_+u}\,,\qquad\qquad V_+= \frac{1}{\kappa_+}e^{\kappa_+v}\,.
\end{equation} 
It follows that
\begin{equation}\label{eq_Kruskal_horizon_IE}
\begin{array}{c} 
    U_+\sim (r-r_+)=0\qquad \text{at} \qquad \mathscr{H}^+_R\,;\qquad\qquad  
    U_-\sim (r-r_-) =0\qquad \text{at} \qquad \mathscr{H}^-_L\,; \\
    V_+\sim (r-r_+)=0\qquad \text{at} \qquad \mathscr{H}^+_L\,;\qquad \qquad 
     V_-\sim (r-r_-)=0\qquad \text{at} \qquad \mathscr{H}^-_R\,.
\end{array}
\end{equation}
The expectation values of of the stress energy tensor expressed in the coordinates regular at the inner horizon read
\begin{subequations}\label{eq:B_state_Kruskal2}
\begin{align}
     \left<B\right|\hat{T}_{U_- U_-}\left|B\right>&=\left(\frac{d u}{d U_-}\right)^2 \left<B\right|\hat{T}_{uu}\left|B\right>=4\left(U_-\right)^{-6}\left<B\right|\hat{T}_{uu}\left|B\right>\,,
     \\
    \left<B\right|\hat{T}_{V_-V_-}\left|B\right>&=\left(\frac{d v}{d V_-}\right)^2 \left<B\right|\hat{T}_{vv}\left|B\right>=4\left(V_-^{-6}\right)\left<B\right|\hat{T}_{vv}\left|B\right>\,,
    \\
    \left<B\right|\hat{T}_{U_- V_-}\left|B\right>&=\frac{d u}{d U_-}\frac{d v}{d V_-} \left<B\right|\hat{T}_{uv}\left|B\right>=4\left(U_-\right)^{-3} \left(V_-\right)^{-3}\left<B\right|\hat{T}_{uv}\left|B\right>\,.
     \end{align}
\end{subequations}
In particular, we are interested in the values near the inner horizon
\begin{subequations}\label{eq:B_state_Kruskal_near_hor}
\begin{align}
     \left<B\right|\hat{T}_{U_- U_-}\left|B\right>&=4\left(U_-\right)^{-6}\left\{\frac{(r_+-r_-)^2}{64\pi Q_4^2(r)}(r-r_-)^4+\right.\\
     
     &\left.\frac{(r_--r_+) \left[r_- (r_--r_+)
   \left(Q_4'(r_-)+2 M r_-^2\right)-2 r_+
   Q_4(r_-)\right]}{32 \pi  r_-
   Q_4(r_-)^3}^5+\mathcal{O}(r-r_-)^6\right\}\,,\nonumber
     \\
    \left<B\right|\hat{T}_{V_-V_-}\left|B\right>&=4\left(V_-\right)^{-6}\left\{\frac{(r_+-r_-)^2}{64\pi Q_4^2(r)}(r-r_-)^4+\right.\\
     
     &\left.\frac{(r_--r_+) \left[r_- (r_--r_+)
   \left(Q_4'(r_-)+2 M r_-^2\right)-2 r_+
   Q_4(r_-)\right]}{32 \pi  r_-
   Q_4(r_-)^3}^5+\mathcal{O}(r-r_-)^6\right\}\,,\nonumber
    \\
    \left<B\right|\hat{T}_{U_- V_-}\left|B\right>&=4\left(U_-\right)^{-3} \left(V_-\right)^{-3}\left[\frac{r_+-r_-}{16\pi Q_4^2(r)}(r-r_-)^4+\mathcal{O}(r-r_-)^5\right]\,.
     \end{align}
\end{subequations}
Note that for the diagonal component we provide the expression to the next to leading order as that will be useful in analysing the stability of the dynamical case. Comparing these expressions with the behavior of the coordinates at horizons provided in Eq.~\eqref{eq_Kruskal_horizon_IE} we see that the $(U_-,U_-)$ component diverges at $\mathscr{H}^-_L$, the $(V_-,V_-)$ component diverges at $\mathscr{H}^-_R$, while the $(U_-,V_-)$ component is everywhere well defined.

\subsubsection{Unruh states (inner and outer)}

As for the non-extremal case, we can compute the expectation values in different states. The Unruh state $\left|U_+\right>$ regular at the outer horizon is obtained in complete analogy with the analysis of the non-extremal case. The Unruh state $\left|U_-\right>$ regular at the inner horizon is obtained by splitting the positive and negative energy modes according to
\begin{equation}\label{eq:Unruh_-_IE}
    \phi=\sum_\omega e^{-i\omega v}a^L_\omega+e^{-i\omega U_-}a^R_\omega+h.c.\,.
\end{equation}
To obtain the expression of the expectation values we need to compute the expression of the Schwarzian derivative
\begin{equation}
    \left\{U_-,u\right\}=\frac{3}{8}u^{-2}\,.%\qquad    \left\{V_-,v\right\}=\frac{3}{8}v^{-2}\,.
\end{equation}
We obtain
\begin{align}
   \left<U_-\right|\hat{T}_{U_- U_-}\left|U_-\right>=&\frac{4(U_-)^{-6}}{96
   \pi  r_-^6 (b+2 M
   r_-)^4} \left[ b^2
   \left(-3 r_-^2+12
   r_- r_+-23
   r_+^2\right)-2 b M
   r_- \left(23
   r_-^2-75 r_-
   r_++80
   r_+^2\right)\nonumber\right.\\
   &\qquad\left.-4 M^2
   r_-^2 \left(23
   r_-^2-69 r_- r_++60 r_+^2\right)\right](r-r_-)^6+(U_-)^{-6}O\left[(r-r_-)^7\right]\,.
\end{align}
Thus this component is finite everywhere. 
On the other hand
\begin{align}
     \left<U_-\right|\hat{T}_{V_- V_-}\left|U_-\right>=&  \left<B\right|\hat{T}_{V_-V_-}\left|B\right>,
    \\
   \left<U_-\right|\hat{T}_{U_- V_-}\left|U_-\right>=& \left<B\right|\hat{T}_{U_- V_-}\left|B\right>\,.
\end{align}
Therefore, the $(V_-,V_-)$ component diverges at $\mathscr{H}_R^-$.
The behaviour at the outer horizons is the same as in the previous cases. So the RSET is divergent there. In particular, in the Boulware state we have
\begin{align}
    &   \left<B\right|\hat{T}_{U_+U_+}\left|B\right>= -\kappa_+^{-2}U^{-2}_+\frac{(r_+-r_-)^6}{192 \pi  Q_4(r_+)^2}\\&
    \left<B\right|\hat{T}_{V_+V_+}\left|B\right>= -\kappa_+^{-2}V^{-2}_+\frac{(r_+-r_-)^6}{192 \pi  Q_4(r_+)^2}    \\&
    \left<B\right|\hat{T}_{V_+U_+}\left|B\right>= -\kappa_+^{-2}V_+U_+\mathcal{O}(r-r_+)\,.
\end{align}
Implying divergences on all four branches of the horizons. On the other hand, the expectation value in the Unruh state $\left|U_+\right>$ 
\begin{align}
&\left<U_+\right|\hat{T}_{U_+ U_+}\left|U_+\right>=\kappa_+^{-2}U^{-2}_+\left[\mathcal{O}(r-r_-)^2\right]\\
%\frac{(r-r_+)^2 (r_--r_+)^4 \left(2(r_--r_+)^2 Q_4'(r_+)^2-Q_4(r_+) (r_--r_+) \left((r_--r_+) Q_4''(r_+)-6 Q_4'(r_+)\right)+6Q_4(r_+)^2\right)}{64 \pi  Q_4(r_+)^4}
     &\left<U_+\right|\hat{T}_{V_+ V_+}\left|U_+\right>=\kappa_+^{-2}V^{-2}_+
\left<B\right|\hat{T}_{vv}\left|B\right>\,\\&
    \left<U_+\right|\hat{T}_{V_+U_+}\left|U_+\right>=
    \left<B\right|\hat{T}_{V_+U_+}\left|B\right>
\end{align}
implying that the state is regular on $\mathscr{H}^+_R$.

\subsection{Summary of the results for eternal geometries}

\begin{figure}[!h]
    \centering
    \includegraphics[width=0.9\linewidth]{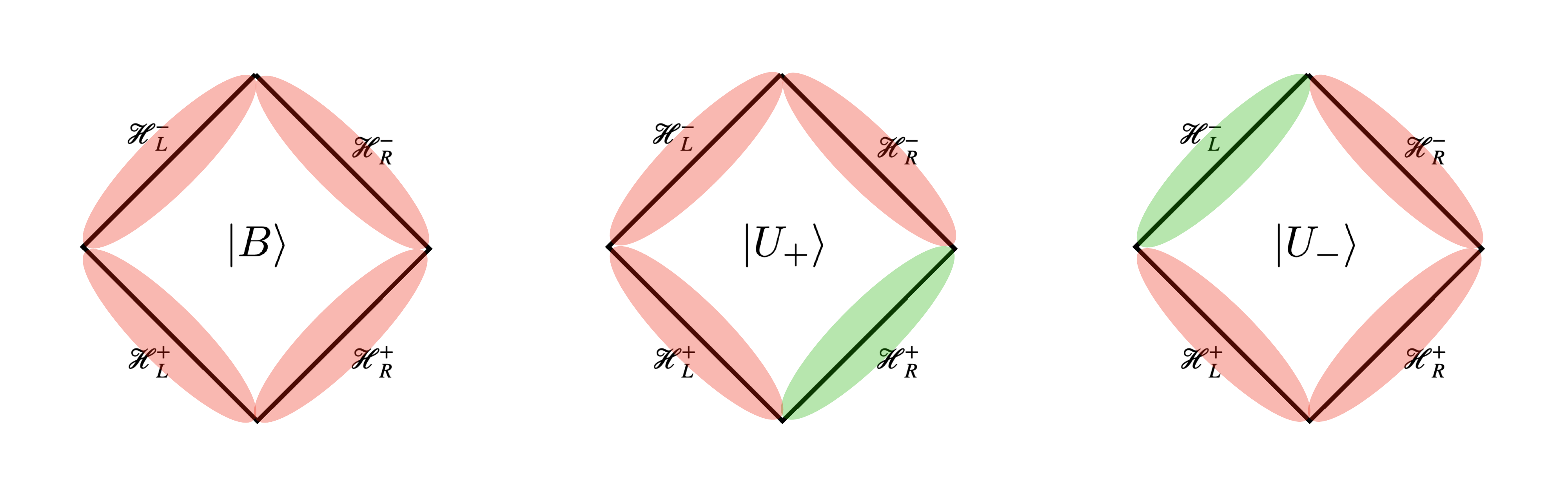}
    \caption{Eternal black holes: Graphical summary of the (lack of) regularity of the different vacuum states defined --- Boulware $|B\rangle$, Unruh $|U_+\rangle$ and inner Unruh $|U_-\rangle$. We do not differentiate among the three static geometries specifically considered (Reissner--Nordstr\"om, Hayward, inner-extremal) as the qualitative behaviour of the RSET is the same. In fact, these considerations apply to any static spacetime with the same horizon structure.}
    \label{fig:summary}
\end{figure}

The qualitative features of the RSET, in particular regarding its divergences, for all the different static spacetimes discussed above. The Boulware state $|B\rangle$ is singular on $\mathscr{H}^+_L$, $\mathscr{H}^+_{R}$, $\mathscr{H}^-_L$ and $\mathscr{H}^-_R$. The Unruh state $|U_+\rangle$ is regular on $\mathscr{H}^+_R$, while the inner Unruh state $|U_-\rangle$ is regular on $\mathscr{H}^-_L$.\footnote{Let us note for completeness that it is possible to define other ``Unruh" states that are regular either on $\mathscr{H}^+_L$ or $\mathscr{H}^-_R$, although we are not considering these here as we are focusing on states that have a clear connection to the dynamical situations analyzed in the second part of the paper.} These features are summarized in Fig.~\ref{fig:summary}, as well as tables below.

%------------------------------------------

\begin{table}[!h]
\begin{tabular}{|c||c|c|c||c|c|c|c|c|c|}
\hline
$|B\rangle$& $\left<\hat{T}_{U_+ U_+}\right>$ & $\left<\hat{T}_{U_+ V_+}\right>$ & $\left<\hat{T}_{V_+ V_+}\right>$ &$\left<\hat{T}_{U_- U_-}\right>$ & $\left<\hat{T}_{U_- V_-}\right>$ & $\left<\hat{T}_{V_- V_-}\right>$ \\
 \hline\hline
 $\mathscr{H}^+_L$  & Finite & Finite  & Divergent & --- & ---& ---\\
 \hline
 $\mathscr{H}^+_R$  & Divergent & Finite & Finite &---&---&---\\
 \hline\hline
 $\mathscr{H}^-_L$  & --- & --- & --- & Divergent & Finite & Finite \\
 \hline
 $\mathscr{H}^-_R$ & --- & --- & --- & Finite & Finite & Divergent  \\
 \hline
\end{tabular}
\caption{Eternal black holes: Behavior of the different components of the renormalized stress-energy tensor in the Boulware state $|B\rangle$ for different branches of the outer and inner horizon. The dashed cells indicate that the coordinates being used do not cover the corresponding branch.\label{tab:rsetcompsB}}
\end{table}

\begin{table}[!h]
\begin{tabular}{|c||c|c|c||c|c|c|c|c|c|}
\hline
   $|U_+\rangle$& $\left<\hat{T}_{U_+ U_+}\right>$ & $\left<\hat{T}_{U_+ V_+}\right>$ & $\left<\hat{T}_{V_+ V_+}\right>$ &$\left<\hat{T}_{U_- U_-}\right>$ & $\left<\hat{T}_{U_- V_-}\right>$ & $\left<\hat{T}_{V_- V_-}\right>$ \\
 \hline\hline
 $\mathscr{H}^+_L$  & Finite & Finite  & Divergent & --- & ---& ---\\
 \hline
 $\mathscr{H}^+_R$  & Finite & Finite & Finite &---&---&---\\
 \hline\hline
 $\mathscr{H}^-_L$  & --- & --- & --- & Divergent & Finite & Finite \\
 \hline
 $\mathscr{H}^-_R$ & --- & --- & --- & Finite & Finite & Divergent  \\
 \hline
\end{tabular}
\caption{Eternal black holes: Behavior of the different components of the renormalized stress-energy tensor in the outer Unruh state $|U_+\rangle$ for different branches of the outer and inner horizon. The dashed cells indicate that the coordinates being used do not cover the corresponding branch.\label{tab:rsetcompsU+}}
\end{table}

\enlargethispage{50pt}

\begin{table}[!h]
\begin{tabular}{|c||c|c|c||c|c|c|c|c|c|}
\hline
   $|U_-\rangle$& $\left<\hat{T}_{U_+ U_+}\right>$ & $\left<\hat{T}_{U_+ V_+}\right>$ & $\left<\hat{T}_{V_+ V_+}\right>$ & $\left<\hat{T}_{U_- U_-}\right>$ & $\left<\hat{T}_{U_- V_-}\right>$ & $\left<\hat{T}_{V_- V_-}\right>$ \\
 \hline\hline
 $\mathscr{H}^+_L$  & Divergent & Finite  & Divergent & --- & ---& ---\\
 \hline
 $\mathscr{H}^+_R$  & Finite & Finite & Finite &----&---&---\\
 \hline\hline
 $\mathscr{H}^-_L$  &--- & --- & --- & Finite & Finite & Finite \\
 \hline
 $\mathscr{H}^-_R$ & --- & --- & --- & Finite & Finite & Divergent  \\
 \hline
\end{tabular}
\caption{Eternal black holes: Behavior of the different components of the renormalized stress-energy tensor in the inner Unruh state $|U_-\rangle$ for different branches of the outer and inner horizon. The dashed cells indicate that the coordinates being used do not cover the corresponding branch.\label{tab:rsetcompsU-}}
\end{table}

%------------------------------------------
\section{Dynamical geometries}\label{sec:dynamical}
\begin{figure}[!h]
    \centering
    \includegraphics[width=0.3\linewidth]{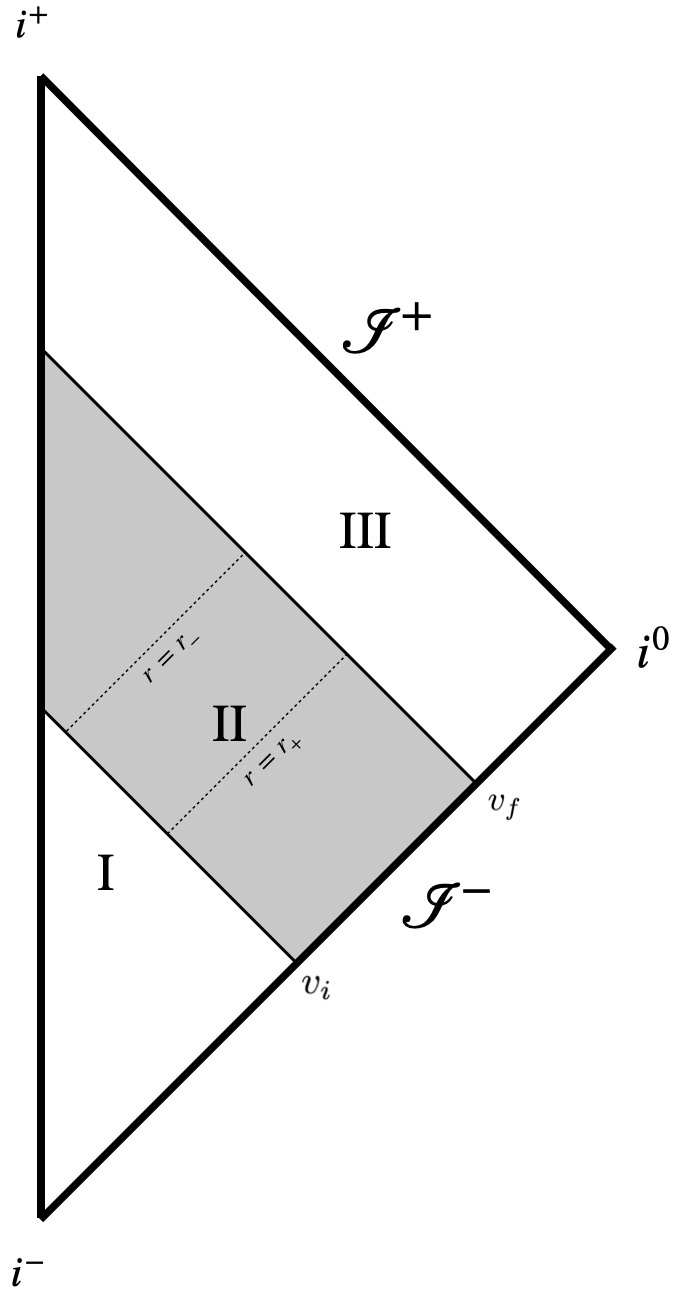}
    \caption{Conformal diagram of the dynamical spacetime under consideration. An incoming shell at $v=v_i$ opens up a trapped region with an inner and outer horizon, region II. The trapped region is then closed at $v=v_f$ by a negative-energy incoming shell, leaving an untrapped region of spacetime, region III.}
    \label{fig:Carter-Penrose}
\end{figure}

%------------------------------------------
We now want to study a quantum field defined on a dynamical geometry that is formed via gravitational collapse, and that evaporates in finite time. We will model the collapsing matter as a null shell.
In the same way, based on the fact that the near-outer-horizon physics of the Hawking flux corresponds to an ingoing negative energy flux, we model the entire evaporation process by an ingoing negative energy null shell.
This is a very crude but remarkably useful approximation. Specifically, the resulting spacetime is described by a regular black hole for an arbitrarily large transient, and is flat in the asymptotic past and flat in the asymptotic future.
Furthermore, all the horizons are dynamical horizons. There is no event horizon nor Cauchy horizon. Therefore, this spacetime captures all the physically relevant aspects we are interested in. The Carter--Penrose diagram is illustrated in Fig.~\ref{fig:Carter-Penrose}.

We consider one continuous coordinate $v$ as the null coordinate on $\mathscr{I}^-$. The other null coordinate is generically discontinuous across the shell, and we shall denote this possibly discontinuous coordinate by $u_\textsc{{i,ii}}$ on the two sides of the collapsing shell.

In $(t,r)$ coordinates the metric in the various regions reads
\begin{equation}
    ds^2_\textsc{{i,iii}}=-dt^2_\textsc{{i,iii}}+dr^2+r^2d\Omega^2
\end{equation}
in the flat regions, and
\begin{equation}
    ds^2_\textsc{{ii}}=-f(r)dt^2_\textsc{{ii}}+f(r)^{-1}dr^2+r^2d\Omega^2
\end{equation}
in the region between the shells.
The time coordinates must be discontinuous across the shells in order to enforce continuity of the induced metric on the shells. In particular, denoting by $R_1$ the radius of the first shell, the induced metric obtained from regions I and II reads
\begin{equation}
    ds^2=-dt_\textsc{{i}}^2+R_1^2d\Omega^2=-f(R_1)dt_\textsc{{ii}}^2+R_1^2d\Omega^2\,,
\end{equation}
implying $dt_\textsc{{i}}^2=f(R_1)dt_\textsc{{ii}}^2$. A similar relation holds between $dt_\textsc{{ii}}$ and $dt_\textsc{{iii}}$.

In the flat regions we define the usual null coordinates as
\begin{equation}
    u_\textsc{{i,iii}}=t_\textsc{{i,iii}}-r\,,
    \qquad
    v=t_\textsc{{i,iii}}+r\,,
\end{equation}
whereas in the interacting region one uses the tortoise coordinate $r^\star$ to define
\begin{equation}
    u_\textsc{{ii}}=t_\textsc{{ii}}-r^\star\,,
    \qquad
    v=t_\textsc{{ii}}+r^\star\,,
\end{equation}
with $dr^\star/dr=f(r)^{-1}$.

The adequate vacuum state to consider is the $|in\rangle$ state, which corresponds to the state with no particles on $\mathscr{I}^-$. The left-going modes are given by
\begin{equation}
    \phi_L\propto e^{-i\omega v}\,,
\end{equation}
while the right-going modes are obtained by reflecting the left-going ones across $r=0$. We have
\begin{equation}\label{eq:in_state}
    \phi=\sum_\omega e^{-i\omega v}a^L_\omega
    +e^{-i\omega u_\textsc{{i}}}a^R_\omega
    +h.c.\, .
\end{equation}

The Fulling--Sweeny--Wald theorem~\cite{Fulling:1978ht} (see also~\cite{FULLING1981243}) states that a quantum state that has a Hadamard singularity structure in an open neighborhood of a Cauchy surface has the same structure everywhere in the Cauchy development of the aforementioned Cauchy surface. Hence, away from the idealized null shells, the expectation values of the stress-energy tensor computed in the $|in\rangle$ state are regular throughout the smooth Cauchy development. This implies that the leading behavior of the $|in\rangle$ state in open neighborhoods of outer and inner horizons is coincident with the behavior of the corresponding Unruh states which we have already defined for the static situation.

In the following, we will show this explicitly for different spacetimes. In particular, understanding the behavior of the $|in\rangle$ state around the inner horizon requires writing explicitly the relations between the null coordinates in regions I and II and defining null coordinates which are regular on the inner horizon. As we discussed explicitly in the static case, the definitions of regular null coordinates on the inner horizon are different depending on the type of geometry considered, and thus the analysis has to be performed on a case-by-case basis. As in the time-independent case, we will first consider a Reissner--Nordstr\"om black hole, a Hayward black hole, and then an inner-extremal regular black hole.

%------------------------------------
\subsection{Transient Reissner--Nordstr\"om spacetime}\label{sec:dynamicalRN}
%------------------------------------

The right-going modes are expressed in terms of the right-going null coordinate in the flat portion of the spacetime. We need to express them as a function of $u_\textsc{{ii}}$, i.e., in terms of the null coordinate of the Reissner--Nordstr\"om portion of the spacetime. In this way, we can perform the Schwarzian derivative in Eq.~\eqref{eq:SET_generic} and obtain the expectation value of the stress-energy tensor in the $|in\rangle$ state.
To this end, we note
\begin{equation}\label{eq:vurel}
    v-u_\textsc{{i}}=2 r\,,
    \qquad
    v-u_\textsc{{ii}}=2r^\star\,,
\end{equation}
where $r^\star$ was defined in Eq.~\eqref{eq:tortoiseRN}.

At the shell, $v=v_0$, we can match these two expressions and obtain
\begin{equation}
%\begin{split}
    v_0-u_\textsc{{ii}}
    =
    v_0-u_\textsc{{i}}
    %&
    +\frac{1}{\kappa_+}
    \log \left|
    \kappa_+\left(\frac{v_0-u_\textsc{{i}}}{2}-r_+\right)
    \right|
    %\\
    %&
    +\frac{1}{\kappa_-}
    \log \left|
    \kappa_-\left(\frac{v_0-u_\textsc{{i}}}{2}-r_-\right)
    \right|\, .
%\end{split}
\end{equation}
Recall that $\kappa_+>0$ and $\kappa_-<0$.

Near the inner horizon, where $v_0-u_\textsc{{i}}\approx 2r_-$, the singular term is the logarithm involving $r-r_-$. Thus
\begin{equation}
     v_0-u_\textsc{{ii}}
     \approx
     \frac{1}{\kappa_-}
     \log \left|
     \kappa_-\left(\frac{v_0-u_\textsc{{i}}}{2}-r_-\right)
     \right|\,,
\end{equation}
and therefore
\begin{equation}
     u_\textsc{{i}}
     =
     u_\textsc{{i}}^{(-)}
     +c_1 e^{-\kappa_-u_\textsc{{ii}}}
     +c_2 e^{-2\kappa_-u_\textsc{{ii}}}
     +\mathcal{O}\!\left(e^{-3\kappa_-u_\textsc{{ii}}}\right)
     =
     u_\textsc{{i}}^{(-)}
     +c_1 e^{|\kappa_-|u_\textsc{{ii}}}
     +c_2 e^{2|\kappa_-|u_\textsc{{ii}}}
     +\mathcal{O}\!\left(e^{3|\kappa_-|u_\textsc{{ii}}}\right)\, .
\end{equation}

Similarly, near the outer horizon, where $v_0-u_\textsc{{i}}\approx 2r_+$, the singular term is the logarithm involving $r-r_+$. Hence
\begin{equation}
     v_0-u_\textsc{{ii}}
     \approx
     \frac{1}{\kappa_+}
     \log \left|
     \kappa_+\left(\frac{v_0-u_\textsc{{i}}}{2}-r_+\right)
     \right|\,,
\end{equation}
and therefore
\begin{equation}
     u_\textsc{{i}}
     =
     u_\textsc{{i}}^{(+)}
     +\tilde c_1 e^{-\kappa_+u_\textsc{{ii}}}
     +\tilde c_2 e^{-2\kappa_+u_\textsc{{ii}}}
     +\mathcal{O}\!\left(e^{-3\kappa_+u_\textsc{{ii}}}\right)\, .
\end{equation}
The values of the constants $c_1,c_2,\tilde c_1,\tilde c_2$ can be computed explicitly, but they do not affect the results of this work.

Hence, at leading order, $u_\textsc{{i}}$ approaches the Kruskal coordinate with surface gravity $\kappa_-$ near the inner horizon, and the Kruskal coordinate with surface gravity $\kappa_+$ near the outer horizon. Therefore, the right-going modes in Eq.~\eqref{eq:in_state} coincide with the inner Unruh modes near the inner horizon, and with the outer Unruh modes near the outer horizon. Thus, the expectation value of the stress-energy tensor in the $|in\rangle$ state is well defined. The explicit computation can be found in the literature; see e.g.~\cite{Balbinot:2023vcm}. Instead of repeating the calculation for this spacetime, we will perform a qualitatively similar, but quantitatively different, calculation for the Hayward spacetime in the section below, as preparation for the later discussion of the inner-extremal case.

%------------------------------------
\subsection{Transient Hayward spacetime}\label{sec:dynamicalH}
%------------------------------------

Following the same steps as before, we have
\begin{equation}
    v-u_\textsc{{i}}=2 r\,,
    \qquad
    v-u_\textsc{{ii}}=2r^\star\,,
\end{equation}
where now $r^\star$ is given by Eq.~\eqref{eq:tortoiseHay}. The leading behavior of $u_\textsc{{i}}$ near the outer and inner horizons is the same as in the previous section. In particular, this implies the following expressions for the corresponding Schwarzian derivatives:
\begin{equation}
    \begin{array}{cc}
   \{u_\textsc{{i}},u_\textsc{{ii}}\}\approx  & -\frac12 \kappa_+^2,
   \qquad\qquad\text{for }\qquad r\approx r_+\,,
   \\[0.5em]
  \{u_\textsc{{i}},u_\textsc{{ii}}\}\approx  & -\frac12 \kappa_-^2,
  \qquad\qquad\text{for }\qquad r\approx r_-\,.
\end{array}
\end{equation}
However, we need to compute the next-to-leading order, as its contribution to the expectation value of the stress-energy tensor does not vanish. A straightforward computation gives
\begin{equation}
    \{u_\textsc{{i}},u_\textsc{{ii}}\}
    =
    -\frac12 \kappa_-^2
    -6\frac{c_2^2}{c_1^2}\kappa_-^2 e^{2|\kappa_-|u_\textsc{{ii}}}
    +\mathcal{O}\!\left(e^{3|\kappa_-|u_\textsc{{ii}}}\right)\, .
\end{equation}
The leading Schwarzian term is the same as in the inner Unruh state, ensuring regularity at the inner horizon. The subleading term in the map $u_\textsc{i}(u_\textsc{ii})$ gives a finite, state-dependent contribution on the horizon.
\subsubsection{The ``in'' state}

With these results for the Schwarzian derivative, we can write down the expectation values of the stress-energy tensor in the $|in\rangle$ state:
\begin{align}
    \left<in\right|\hat{T}_{U_- U_-}\left|in\right>
    =
    \kappa_-^{-2}U^{-2}_-
    \Bigg[
    &\frac{
    r_-^7+4r_-^6 r_+-12r_-^5 r_+^2-50r_-^4 r_+^3-11r_-^3 r_+^4
    +66r_-^2 r_+^5+22r_- r_+^6-20r_+^7
    }{
    32 \pi r_-^3
    \left(r_-^2+r_-r_++r_+^2\right)^4
    }
    (r-r_-)^2\nonumber\\
    &+\mathcal{O}\!\left[(r-r_-)^3\right]
    \Bigg]+U^{-2}_-
    \Bigg[
    \frac{1}{4\pi}\frac{c_2^2}{c_1^2}
    e^{2|\kappa_-|u_\textsc{{ii}}}
    +\mathcal{O}\!\left(e^{3|\kappa_-|u_\textsc{{ii}}}\right)
    \Bigg]\, .
\end{align}
We also have
\begin{align}
     \left<in\right|\hat{T}_{V_- V_-}\left|in\right>
     &=
     \kappa_-^{-2}V^{-2}_-
     \left\{
     -\frac{\left(r_-^2+r_- r_+-2r_+^2\right)^2}{192 \pi r_-^2 \left(r_-^2+r_-r_++r_+^2\right)^2}
     +\mathcal{O}\!\left[(r-r_-)^2\right]
     \right\}\, .
\end{align}
{We can see that for any finite lifetime of the trapped region $V_-$ remains finite and non-zero, so this component is finite. The factor $V_-^{-2}$ nevertheless displays the approach to the would-be Cauchy-horizon divergence in the long-lifetime limit, $|v_f-v_i|\to\infty$.}

Comparing this result with the one in Eq.~\eqref{eq:Hayward_Uminus}, we can see that at the lowest order the expectation value in the $|in\rangle$ state is equal to the expectation value in the $|U_-\rangle$ state, with differences arising only at next-to-leading order
\begin{equation}
    \left<in\right|\hat T_{U_-U_-}\left|in\right>
    =
    \left<U_-\right|\hat T_{U_-U_-}\left|U_-\right>
    +\mathcal{O}(r-r_-)+\Delta_-\,,
\end{equation}
where $ \Delta_-$ is a finite and state dependent contribution coming from the expansion of the Schwarzian.

Similarly, close to the outer horizon,
\begin{align}
    \left<in\right|\hat{T}_{U_+ U_+}\left|in\right>
    =
    \kappa_+^{-2}U^{-2}_+
    \Bigg[
    &\frac{
    r_+^7+4r_+^6 r_--12r_+^5 r_-^2-50r_+^4 r_-^3-11r_+^3 r_-^4
    +66r_+^2 r_-^5+22r_+ r_-^6-20r_-^7
    }{
    32 \pi r_+^3
    \left(r_+^2+r_+r_-+r_-^2\right)^4
    }
    (r-r_+)^2\nonumber\\
    &+\mathcal{O}\!\left[(r-r_+)^3\right]
    \Bigg]\, ,
\end{align}
which, as for the inner horizon, satisfies
\begin{equation}
    \left<in\right|\hat{T}_{U_+ U_+}\left|in\right>
    =
    \left<U_+\right|\hat{T}_{U_+ U_+}\left|U_+\right>
    +\mathcal{O}(r-r_+)+\Delta_+\, ,
\end{equation}
{where $ \Delta_+$ is a finite and state dependent contribution coming from the expansion of the Schwarzian.}

\subsubsection{Physical interpretation}

Before moving to the inner-extremal spacetime, let us pause and discuss the results of this section.
We have seen that the expectation values of the stress-energy tensor in the $|in\rangle$ state are finite everywhere.
This result does not necessarily imply that the backreaction of the quantum effects would be small. Classical mass-inflation computations show that counter-streaming fluxes of energy lead to an exponentially large backreaction. The diagonal components of the stress-energy tensor correspond to two fluxes of energy which we expect would give rise to an instability once the backreaction is taken into account.

Finally, if the lifetime of the evaporation is very large, we expect that we should recover the results of Sec.~\ref{sec:eternal} and obtain a divergence as we approach the right sector of the Cauchy horizon. At first sight, we do not recover this limit as none of the expectation values depends explicitly on $v$. However, we have worked in the $(u_\textsc{{ii}},v)$ coordinates that are well defined for the dynamical geometry but not in the eternal limit, in which a right sector of the Cauchy horizon would form. Rewriting the expectation values in the $(u_\textsc{{ii}},V_-)$ coordinates, we obtain
\begin{equation}
    \left<in\right|\hat{T}_{V_- V_-}\left|in\right>
    =
    \left(\frac{\partial v}{\partial V_-}\right)^2
    \left<in\right|\hat{T}_{vv}\left|in\right>
    =
    \kappa_-^{-2}V_-^{-2}
    \left<in\right|\hat{T}_{vv}\left|in\right>
    \propto
    e^{2|\kappa_-|v}
    \left<in\right|\hat{T}_{vv}\left|in\right>\, .
\end{equation}
Thus, while an ingoing null observer measures a finite flux, an outgoing null observer would measure an exponentially growing stress-energy tensor in the long-lifetime limit.

%------------------------------------
\subsection{Transient inner-extremal spacetime}\label{sec:dynamicalIE}
%------------------------------------

The relation between the $u_\textsc{{ii}}$ and $u_\textsc{{i}}$ coordinates can be found through Eq.~\eqref{eq:vurel}, with $r^\star$ defined in Eq.~\eqref{eq:tort_Inner_extr}. Combining these relations with Eq.~\eqref{eq:tort_Inner_extr}, and setting
\begin{equation}
    x=\frac{v_0-u_\textsc{{i}}}{2}-r_-\,,
\end{equation}
we obtain
\begin{equation}
%\begin{split}
    u_\textsc{{ii}}
    =
    u_\textsc{{i}}
    -\frac{2A}{\left(\frac{v_0-u_\textsc{{i}}}{2}-r_-\right)^2}
    -\frac{2B}{\left(\frac{v_0-u_\textsc{{i}}}{2}-r_-\right)}
    -2C\log\left|
    \kappa_+\left(\frac{v_0-u_\textsc{{i}}}{2}-r_-\right)
    \right|
    -\frac{1}{\kappa_+}
    \log \left|
    \kappa_+\left(\frac{v_0-u_\textsc{{i}}}{2}-r_+\right)
    \right|\, .
%\end{split}
\end{equation}
The constants $A$, $B$, and $C$ are given in Eqs.~\eqref{eq:A-B} and~\eqref{eq:C}.

Let us now study the regularity of the stress-energy tensor at the inner horizon. Near the inner horizon we have
\begin{equation}\label{eq:u_I-inhor}
    u_\textsc{{ii}}
    =
    -\frac{2A}{\left(\frac{v_0-u_\textsc{{i}}}{2}-r_-\right)^2}
    -\frac{2B}{\left(\frac{v_0-u_\textsc{{i}}}{2}-r_-\right)}
    +\mathcal{O}\!\left[
    \log\left|\frac{v_0-u_\textsc{{i}}}{2}-r_-\right|
    \right]\, .
\end{equation}
This implies
\begin{equation}
     u_\textsc{{i}}
     =
     u_\textsc{{i}}^{(-)}
     +\mathcal{O}\!\left(u_\textsc{{ii}}^{-1/2}\right)\, .
\end{equation}
We can see that, once again, the coordinate $u_\textsc{{i}}$ approaches the regular inner-horizon coordinate $U_-$ obtained in Eq.~\eqref{eq:U_-innerextr}.

\subsubsection{The ``in'' state}

The expectation value of the stress-energy tensor is
\begin{equation}
    \left<in\right|T_{u_\textsc{{ii}}u_\textsc{{ii}}}\left|in\right>
    =
    -\frac{1}{192\pi}
    \left[
    f'(r)^2-2f(r)f''(r)
    \right]
    -\frac{1}{24\pi}
    \left\{u_\textsc{{i}},u_\textsc{{ii}}\right\}\, .
\end{equation}
Let us remind the reader that, in order for this expectation value to be regular at the inner horizon, it must be regular in a set of coordinates $\{U_-,v\}$ well defined there. Furthermore,
\begin{equation}\label{eq:SET_inextr_Kruskal}
    \left<in\right|T_{U_-U_-}\left|in\right>
    =
    \left<in\right|T_{u_\textsc{{ii}}u_\textsc{{ii}}}\left|in\right>
    \left(\frac{du_\textsc{{ii}}}{dU_-}\right)^2
    \propto
    \left<in\right|T_{u_\textsc{{ii}}u_\textsc{{ii}}}\left|in\right>
    \left[
    (r-r_-)^{-6}
    +\mathcal{O}\!\left((r-r_-)^{-5}\right)
    \right]\, .
\end{equation}
Therefore, $\left<in\right|T_{u_\textsc{{ii}}u_\textsc{{ii}}}\left|in\right>$ must vanish at least as rapidly as $(r-r_-)^6$.

In order to verify this explicitly, we expand the two contributions to $\left<in\right|T_{u_\textsc{{ii}}u_\textsc{{ii}}}\left|in\right>$ near the inner horizon. We have
\begin{align}\label{eq:SET_in_I}
   -\frac{1}{192\pi}
   \left[
   f'(r)^2-2f(r)f''(r)
   \right]
   &=
   \frac{3\left(r_+-r_-\right)^2}{192\pi Q_4(r_-)^2} \left(r-r_-\right)^4
   \nonumber\\
   &\quad
   +\frac{(r_--r_+) \left[ (r_+-r_-)Q_4'(r_-)+Q_4(r_-) \right]}{16 \pi Q_4(r_-)^3} \left(r-r_-\right)^5
   \nonumber\\
   &\quad
   +\mathcal{O}\!\left((r-r_-)^6\right)\, .
\end{align}
The expression of the Schwarzian derivative is involved. However, we can easily compute its value near the horizon. Near the inner horizon, Eq.~\eqref{eq:u_I-inhor} gives
\begin{equation}
    \left\{u_\textsc{{i}},u_\textsc{{ii}}\right\}
    =
    \frac{3}{8}u_\textsc{{ii}}^{-2}
    +\mathcal{O}\!\left(u_\textsc{{ii}}^{-3}\right)\, .
\end{equation}
Using
\begin{equation}
    u_\textsc{{ii}}
    \approx
    -\frac{2A}{(r-r_-)^2}
    -\frac{2B}{r-r_-}\, ,
\end{equation}
we obtain
\begin{equation}
    -\frac{1}{24\pi}
    \left\{u_\textsc{{i}},u_\textsc{{ii}}\right\}
    =
    -\frac{1}{256\pi A^2}(r-r_-)^4
    +\frac{B}{128\pi A^3}(r-r_-)^5
    +\mathcal{O}\!\left[(r-r_-)^6\right]\, .
\end{equation}
Combining this result with Eq.~\eqref{eq:SET_in_I} and replacing the values of the constants $A$ and $B$ given in Eq.~\eqref{eq:A-B}, we obtain
\begin{equation}
  \left<in\right|T_{u_\textsc{{ii}}u_\textsc{{ii}}}\left|in\right>
  =
  \mathcal{O}\!\left[(r-r_-)^6\right]\, .
\end{equation}
Thus, from Eq.~\eqref{eq:SET_inextr_Kruskal}, we conclude that the expectation value of the stress-energy tensor is regular in the regular inner-horizon coordinates, showing that there is no divergence at the inner horizon.

Similarly, we can study the regularity at the outer horizon. Near the outer horizon we have
\begin{equation}\label{u_I(u_II)}
    u_\textsc{{ii}}
    \approx
    -\frac{1}{\kappa_+}
    \log \left|
    \kappa_+\left(\frac{v_0-u_\textsc{{i}}}{2}-r_+\right)
    \right|\, .
\end{equation}
Therefore, the relation between $u_\textsc{{ii}}$ and $u_\textsc{{i}}$ is the same relation that there is between $u_\textsc{{ii}}$ and the Kruskal coordinate $U_+$ defined at the outer horizon, implying that the $|in\rangle$ state approximates the Unruh state with temperature $\kappa_+$ at the outer horizon.

We can check this explicitly. The Schwarzian derivative close to the outer horizon reads
\begin{equation}
     \{u_\textsc{{i}},u_\textsc{{ii}}\}
     =
     -\frac12 \kappa_+^2
     +\mathcal{O}(r-r_+)\, .
\end{equation}
This implies
\begin{equation}
    \left<in\right|\hat{T}_{U_+ U_+}\left|in\right>
    =
    \left<U_+\right|\hat{T}_{U_+ U_+}\left|U_+\right>
    +\mathcal{O}(r-r_+)\, ,
\end{equation}
which is regular at the outer horizon as seen in the previous sections.

\subsubsection{Physical interpretation}

The expectation values of the various components of the stress-energy tensor are finite everywhere in the $|in\rangle$ state for the inner-extremal geometry. This is the same result obtained for the non-extremal geometries. However, the interpretation of the result is different. While finite fluxes of energy are expected to lead to a large instability for non-extremal geometries, we expect the backreaction to stay small for inner-extremal geometries, as such geometries are not affected by the mass-inflation instability.

Another difference concerns the way the divergence in the eternal limit is approached. As for the non-extremal case, we need to consider the expectation values in the coordinates $(u_\textsc{{ii}},V_-)$. For an inner-extremal horizon we have $V_-\propto v^{-1/2}$, and therefore
\begin{equation}
    \left(\frac{\partial v}{\partial V_-}\right)^2
    \propto
    V_-^{-6}
    \propto
    v^3\, .
\end{equation}
Thus
\begin{equation}
    \left<in\right|\hat{T}_{V_- V_-}\left|in\right>
    =
    \left(\frac{\partial v}{\partial V_-}\right)^2
    \left<in\right|\hat{T}_{vv}\left|in\right>
    \propto
    v^3
    \left<in\right|\hat{T}_{vv}\left|in\right>\, .
\end{equation}
Near the inner-extremal horizon, $\left<in\right|\hat{T}_{vv}\left|in\right>\propto (r-r_-)^4$, while along the right inner horizon $v\propto (r-r_-)^{-2}$. Hence $\left<in\right|\hat{T}_{vv}\left|in\right>\propto v^{-2}$, and therefore
\begin{equation}
    \left<in\right|\hat{T}_{V_- V_-}\left|in\right>
    \propto v\, .
\end{equation}
Thus, in the inner-extremal case, the divergence in the eternal limit is approached only as a power law, rather than exponentially in time.

\subsection{Summary of the results for dynamical geometries}

The situation for dynamical geometries (at least the sub-class of dynamical geometries we are considering) is summarized in Table~\ref{tab:rset_instate} below.

\begin{table}[!h]
\begin{tabular}{|c||c|c|c||c|c|c|c|c|c|}
\hline
   $|in\rangle$& $\left<\hat{T}_{U_+ U_+}\right>$ & $\left<\hat{T}_{U_+ V_+}\right>$ & $\left<\hat{T}_{V_+ V_+}\right>$ &$\left<\hat{T}_{U_- U_-}\right>$ & $\left<\hat{T}_{U_- V_-}\right>$ & $\left<\hat{T}_{V_- V_-}\right>$ \\
 \hline\hline
 $r=r_+$  & Finite & Finite  & Finite & --- & ---& ---\\
 \hline
 $r=r_-$  & --- & --- & --- & Finite &Finite&Finite\\
 \hline
\end{tabular}
\caption{Dynamical black holes: Behavior of the different components of the renormalized stress-energy tensor in the $in$ state $|in\rangle$ at the outer and inner horizons. The dashed cells indicate that the coordinates being used do not cover the corresponding horizon. The entries above qualitatively apply for all of the considered geometries with the caveat that the $\left<\hat{T}_{V_- V_-}\right>$ component grows exponentially with the trapped region lifetime (as measured in advanced time) in the Reissner--Nordstr\"om and Hayward geometry, but only linearly in $v$ in the inner-extremal one.
\label{tab:rset_instate}}
\end{table}

%=======================================================
\section{Conclusions}\label{sec:conclusions}
%=======================================================

In this work we have investigated the semiclassical regularity of spacetimes containing compact trapped regions with inner horizons, with particular emphasis on the distinction between eternal geometries, where inner horizons are Cauchy horizons, and dynamical geometries, describing formation and evaporation of  a trapped region in finite time. Working in the s-wave Polyakov approximation, we have computed the renormalized stress-energy tensor for a set of representative geometries, including Reissner--Nordström, Hayward, and inner-extremal black holes.

For eternal geometries, our analysis confirms the standard obstruction to constructing a quantum state that is regular on all branches of the horizon simultaneously. The Boulware state is singular at both outer and inner horizons, while Unruh-like states can be made regular on a selected branch of either the outer or the inner horizon, but not on the full horizon structure. In particular, as already noted in the extant literature \cite{McMaken:2023uue}, the presence of an inner-extremal horizon does not by itself remove this obstruction in the stationary setting: Although the vanishing of the inner surface gravity eliminates the classical mass-inflation mechanism, the RSET still displays singular behaviour in the corresponding eternal spacetime. From the point of view of strict regularity of stationary quantum states, inner-extremal geometries are therefore not qualitatively better than their non-extremal counterparts.

The situation changes substantially once one considers dynamical geometries describing the formation and disappearance of a compact trapped region in finite time. In this case there is no Cauchy horizon, and the natural quantum state is the $in$-state defined on past null infinity. Consistently with the Fulling--Sweeny--Wald theorem, the RSET in this state remains finite everywhere. Near the outer horizon, the $in$-state reproduces asymptotically the local behaviour of the usual Unruh state. Near the inner horizon, it instead reproduces the appropriate inner-Unruh behaviour, ensuring regularity also there. 

Thus, the divergences found in eternal geometries are not local, instantaneous obstructions associated with the mere presence of an inner trapping horizon; rather, they arise in the limiting case in which the trapped region persists indefinitely and a Cauchy horizon is formed. As such they are unphysical: they are at best red-herrings saying that the eternal spacetimes --- characterized by Cauchy and event horizons ---cannot arise in any physically plausible evolution of a dynamical geometry.\footnote{It is perhaps worth emphasizing that this point of view was also advocated by Stephen Hawking in his later years~\cite{Hawking:2014tga}.}

This finiteness in the $in$-state, however, should not be confused with semiclassical stability. For non-extremal inner horizons, the relevant RSET components exhibit an accumulation effect which, when expressed in coordinates adapted to the would-be Cauchy horizon, grows exponentially with time, with a rate controlled by the inner-horizon surface gravity. Hence, for sufficiently long-lived trapped regions, the dynamical result approaches the familiar stationary divergence. Although the stress-energy tensor remains finite at any finite time, its growth suggests that semiclassical backreaction can become large and may destabilize the inner region, in close analogy with the classical mass-inflation instability. 

This is exactly the sort of behavior described in~\cite{Barcelo:2020mjw,Barcelo:2022gii} and found recently in numerical simulations \cite{Barenboim:2025ckx, Boyanov:2025otp} and analytical toy models \cite{Arrechea:2026-in-preparation}. {The calculation of such backreaction remains within the realm of general relativity for the Reissner--Nordstr\"om spacetime, while the Hayward and inner-extremal spacetimes require a separate treatment involving modifications of the Einstein equations, with the formalism discussed in~\cite{Carballo-Rubio:2025ntd} providing a framework in which backreaction can be computed for the latter two cases.

As in the inner-extremal case the inner-horizon surface gravity vanishes, the exponential amplification is replaced by a much milder power-law growth. In the explicit model analysed here, the would-be eternal divergence is approached only linearly in the appropriate inner-horizon coordinate, rather than exponentially. This mirrors the classical result that inner-extremal geometries evade mass inflation \cite{Carballo-Rubio:2022kad}, and suggests that they may be natural candidates for long-lived, semiclassically meta-stable black-hole interiors. Establishing this conclusion definitively will require solving the semiclassical backreaction problem rather than treating the RSET on a fixed background. Nevertheless, the results presented here indicate that inner extremality may provide the relevant geometric mechanism for simultaneously avoiding Cauchy-horizon singularities in finite-lifetime spacetimes and suppressing the accumulation of semiclassical energy near the inner horizon.

\begin{acknowledgments}

The authors wish to thank J.~Arrechea and M.~Spadafora for illuminating discussions.

\end{acknowledgments}
%========================================================
\bibliographystyle{utphys}
%========================================================
\bibliography{refs}
%========================================================
\bigskip
\hrule\hrule\hrule
%=========================================================
\end{document}